\documentclass[12pt]{iopart}

\usepackage{iopams}
\usepackage{eucal}
\usepackage{graphicx}

\newcommand{\lp}{\left(}
\newcommand{\rp}{\right)}
\newcommand\DC{\Delta_{\mathrm{C}}}
\newcommand\DA{\Delta_{\mathrm{A}}}
\newcommand\Liou{{\mathcal L}}
\newcommand{\comm}[2]{\left[#1,#2\right]}
\newcommand{\ket}[1]{\left\vert #1\right\rangle}
\newcommand{\bra}[1]{\left\langle #1\right\vert}
\newcommand{\braket}[2]{\left\langle #1\left|\vphantom{{#1}{#2}}\right.#2\right\rangle}
\newcommand\avr[1]{\left\langle #1 \right\rangle}
\newcommand\eex[1]{\cdot10^{#1}}
\newcommand{\abs}[1]{\left\vert #1 \right\vert}

\begin{document}
\title{Microscopic physics of quantum self-organisation of optical
lattices in cavities}

\author{Andr\'as Vukics, Christoph Maschler, and Helmut Ritsch}
\address{Institut f\"ur theoretische Physik, Universit\"at
  Innsbruck, Technikerstr.~25, A-6020 Innsbruck, Austria}

\begin{abstract}
We study quantum particles at zero temperature in an optical lattice
coupled to a resonant cavity mode. The cavity field substantially
modifies the particle dynamics in the lattice, and for strong
particle-field coupling leads to a quantum phase with only every
second site occupied. We study the growth of this new order out of a
homogeneous initial distribution for few particles as the microscopic
physics underlying a quantum phase transition. Simulations reveal that
the growth dynamics crucially depends on the initial quantum many-body
state of the particles and can be monitored via the cavity
fluorescence. Studying the relaxation time of the ordering reveals
inhibited tunnelling, which indicates that the effective mass of the
particles is increased by the interaction with the cavity
field. However, the relaxation becomes very quick for large coupling.
\end{abstract}

\maketitle

\section{Introduction}
Ultracold atoms in an optical lattice formed by a far detuned laser
field constitute an ideal system to study quantum phase transitions,
ie phase transitions at zero temperature \cite{sachdev}. In the most
prominent example first predicted theoretically \cite{jaksch98} and
confirmed experimentally \cite{greiner02} it was found that the
particle ground state changes from a superfluid state where all atoms
are delocalised to a perfectly ordered Mott insulator state for
increasing lattice depth. More complex phases as supersolids,
etc.~were predicted if long range interactions or mixed species setups
\cite{buechler03} are used, but these are harder to realize and
measure experimentally. While the final states are well understood,
these phase transitions require the buildup or decay of long range
correlations, the mechanism and time scale of which is not fully
understood.

In a parallel development a dynamical transition to a self-organised
phase in optical lattices was found for classical particles, when the
lattice is placed inside an optical resonator. It originates from
interference of the resonantly enhanced light field scattered by the
atoms into the cavity mode with the lattice light itself and leads to
a preferred occupation of every second site \cite{domokos02b}.  For a
finite temperature cloud thermal density fluctuations are amplified
and lead to a runaway self-organisation by feedback from the cavity
field. However, for a BEC (\(T\approx0\)) the initial density is
perfectly homogeneous and only quantum fluctuations which go beyond a
mean field description of the cold gas can start the self-organising
process when the cavity interaction is switched on. Tunnelling results
in a dynamical change of the atomic phase at \(T=0\) which gets
irreversible only if cavity decay is included. In this work we study
this \emph{quantum} dynamics on the microscopic level and show how it
depends on the precise quantum properties of the initial atomic and
field state beyond any mean field density.

The paper is also intended to show the limitations of the effective
Bose-Hubbard type model developed in \cite{maschler05a} for
atom-cavity systems. We demonstrate that in the regime of moderate
coupling the simple Bose-Hubbard approach reproduces very well the
results of a full Monte Carlo wave-function simulation, while it
breaks down in the regime of stronger coupling. Even in this regime,
however, it predicts the steady state surprisingly well, whereas the
relaxation time to this state is predicted wrong. We show that the
relaxation time of the system exhibits a highly non-trivial
behaviour. In the regime of moderate atom-cavity coupling the
relaxation time is composed of the timescale of photon counts and that
of tunnelling. The combination of these two time scales leads to a
minimum behaviour in the relaxation time, while for stronger coupling
the relaxation becomes very quick, a behaviour observed in the full
simulations but not reproducible with the Bose-Hubbard approach.

We first describe our system, the model, and solution methods
applied. Afterwards, we go on investigating the dynamics of a single
atom in the system. Although self-organisation cannot be defined in
this case, we demonstrate that already here the relaxation time
exhibits the same behaviour we find later for two atoms, and here it
is easier to give a qualitative picture of this behaviour. We finally
turn to the case of two atoms and show how the increasing coupling
results in a transition from a \(T=0\) homogeneous initial condition
into a self-organised configuration in steady state.

\section{System, models}
\begin{figure}
\centering
\includegraphics[width=3in]{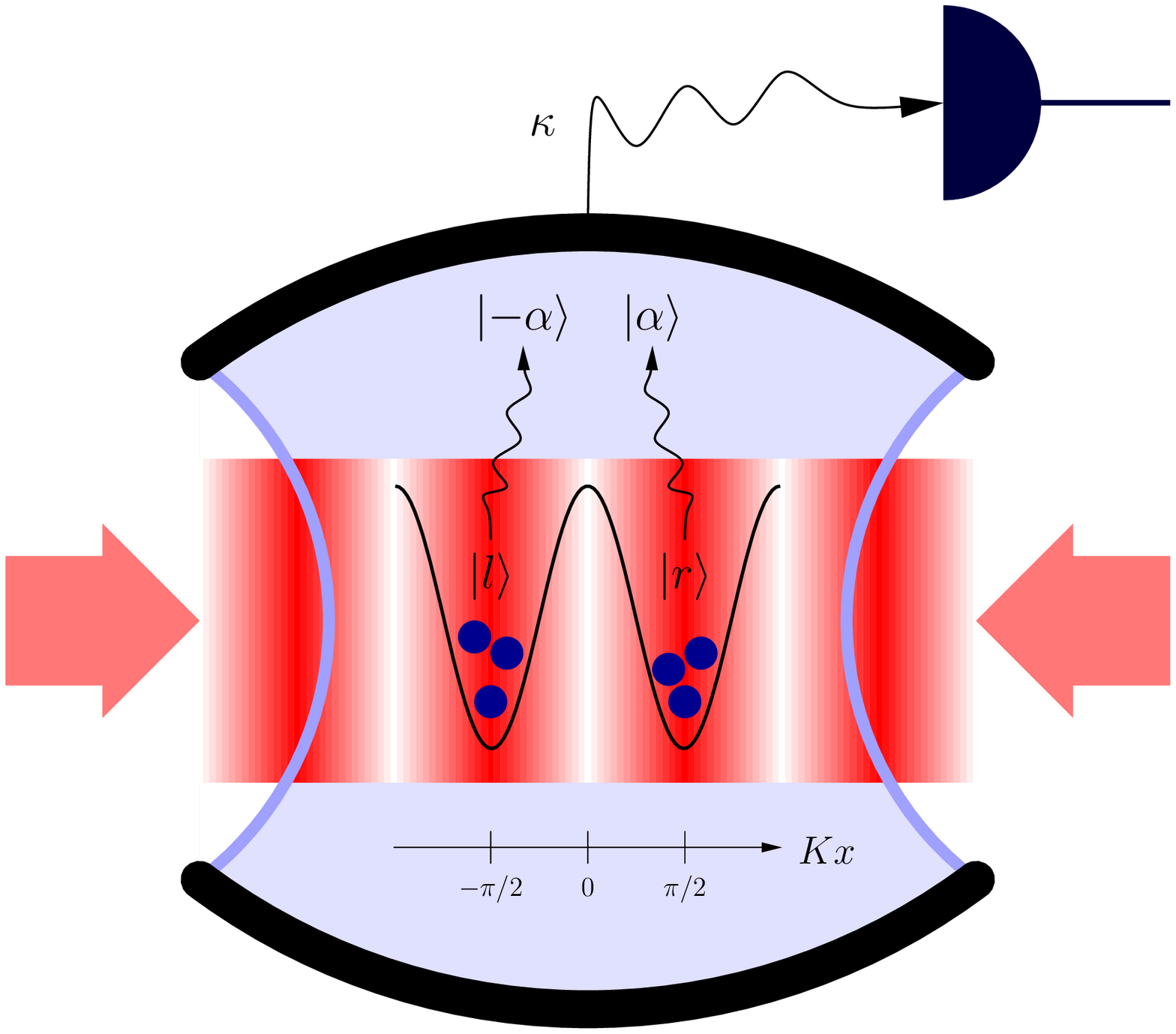}
\caption{Scheme of the system consisting of a one dimensional optical
  lattice with lattice axis \(x\) and a cavity sustaining a single EM
  mode aligned orthogonally.}
\label{fig:Scheme}
\end{figure}

The proposed setup is depicted in \Fref{fig:Scheme}. It consists of a
one dimensional optical lattice within a cavity sustaining a single
mode with its axis aligned orthogonally to the lattice axis --- such
systems have been studied in diverse theoretical contexts before
\cite{domokos04,zippilli05a,metz06,murr06}, and is available
experimentally \cite{nagorny03,nussmann05a}. We assume that the cavity
mode function is constant along the lattice direction, still, as we
will see below, it modifies significantly the dynamics of atoms in the
lattice.

A standard quantum optical model for the system with a single atom
moving in one dimension along the lattice axis is obtained by
adiabatic elimination of the atomic excited state(s). This is
justified in the regime where the driving --- in our case, the laser
generating the lattice (pump) --- is far detuned from the atomic
transition frequency \cite{domokos03,vukics05b}. The Hamiltonian for a
single atom and the cavity then reads (\(\hbar=1\)):
\numparts
\begin{equation}
\label{eq:Ham}
\fl H=\frac{p^2}{2\mu}+V_0\,\sin^2(Kx)
-\lp\DC-U_0\rp a^\dag a+
\mathrm{sign}(U_0)\sqrt{ U_0\,V_0}\,\sin(Kx)\,\lp a^\dag+a\rp.
\end{equation}
Here \(x\), \(p\), and \(\mu\) are the atomic position and momentum
operators and the mass, respectively; \(a\) is the cavity field
operator, \(\DC=\omega-\omega_{\mathrm{C}}\) is the cavity detuning
--- \(\omega\) is the laser, \(\omega_{\mathrm{C}}\) is the cavity
frequency ---, and \(K\) is the lattice field wave number. The first
two terms describe the atomic motion in the lattice, which appears as a
classical potential after the elimination of the atomic internal
dynamics. The potential depth for a two-level atom \( V_0=\eta^2/\DA\)
where \(\eta\) is the pump Rabi frequency and
\(\DA=\omega-\omega_{\mathrm{A}}\), \(\omega_{\mathrm{A}}\) is the
atomic transition frequency. The third term is the free Hamiltonian of
the cavity mode with the detuning shifted by \(U_0=g^2/\DA\) where
\(g\) is the atom-cavity interaction coupling constant. The last term
describes the atom-mode interaction which stems from stimulated
absorption from the pump followed by stimulated emission into the
cavity mode and the reverse process. Atomic spontaneous emission is
strongly suppressed due to the large atom-pump detuning and therefore
neglected. The cavity mode is, however, coupled to the surrounding EM
modes, resulting in the decay of cavity photons (escape through the
mirrors). The process is described by the Liouvillean dynamics
\cite{gardiner}
\begin{equation}
\label{eq:Liou}
\Liou\rho=\kappa\lp2a\rho a^\dag-\comm{a^\dag a}{\rho}_+\rp.
\end{equation}
\endnumparts

Note that model (\ref{eq:Ham}-\ref{eq:Liou}) is not specific to a two-level
atom but generally applicable to linearly polarisable particles, atoms
and molecules alike. In this case the parameter \(U_0\) is
proportional to the susceptibility of the particles
\cite{vukics05b}. In the following we shall hence speak about
\emph{particles} without any further specification.

As described theoretically \cite{domokos02b,asboth05} and observed
experimentally \cite{black03} the model (\ref{eq:Ham}-\ref{eq:Liou})
features a phase transition termed self-organisation for a finite
temperature classical gas. As this occurs for red-detuned driving, ie
high field seeking particles with \(U_0,\,\DC<0\) we will restrict
ourselves to this case. Self-organisation can be qualitatively
understood as follows: The system has three steady-state
configurations in a mean field description: (i) an ``unorganised''
configuration where the particles are equally distributed at all
lattice sites and scatter no light into the cavity due to destructive
interference and (ii) two ``organised'' configurations in which the
particles occupy either only the even or only the odd sites of the
lattice and scatter superradiantly into the cavity. In the latter case
the last term in the Hamiltonian (\ref{eq:Ham}) deepens the lattice
potential \emph{at the positions of the particles}, so that they are
\emph{self-trapped} or ``self-organised''.

Configurations (ii) have lower energy and entropy (lower
symmetry) than configuration (i). At a given temperature the system
chooses between configuration (i) and one of configurations (ii) so as
to minimise the free energy. Lowering the temperature results in a
phase transition when the symmetry of configuration (i) is
spontaneously broken into the lower symmetry of one of configurations
(ii).\footnote{Note that self-organisation is a phase transition
without any direct particle-particle interaction: only an effective
interaction exists generated by the single cavity mode with which all
particles interact.}

The above qualitative picture is modified by the fact that the system
never reaches thermal equilibrium with some external heat bath as
energy is continuously flowing through the system from the pump via
scattering on the particles into the cavity field and then out via the
cavity loss channel. Self-organisation is therefore a dynamical phase
transition for which the above mentioned configurations are
steady-state patterns. In steady state the particles have a momentum
distribution determined by the cavity field fluctuations, which, in
most cases of physical interest, resembles very much a thermal
distribution \cite{vukics05a}. In this sense it is justified to speak
about an effective temperature of the particles and use the picture of
an equilibrium phase transition as we did above.

Let us now turn to the case of zero temperature and envisage a fixed
number of classical point particles at each lattice site. In contrast
to above, no matter whether we are above or below the threshold, no
dynamics will arise because a homogeneous gas scatters no field into
the cavity due to destructive interference. If no photons are present
initially, such a classical gas cannot break the symmetry and is
unable to escape the initial homogeneous configuration (i).

In the following we show that this is quite different for a quantum
Bose gas at \(T=0\), which can be prepared by loading a BEC into the
lattice \cite{ottl06}. Interestingly in this case self-organisation
and the superradiant build-up of the cavity photon number starts
immediately, but depends crucially on the quantum fluctuations of the
gas at \(T=0\). Hence both for a Mott insulator and superfluid state
(BEC) as initial condition, quantum fluctuations and the possibility
of tunnelling between lattice sites immediately start
self-organisation. This is combined with an intricate
measurement-induced dynamics related to the information gained via the
dissipation channel of the photon-loss.

Let us emphasise that there are two main differences as compared to
the above-described classical self-organisation: the initial
temperature of the particles is zero and the particles in the lattice
are confined strongly enough so that hopping between the lattice sites
is due solely to tunnelling. Redistribution thus is a coherent quantum
process and requires no direct inter-particle interaction. 

We are using two approaches to the problem. The first one is the
direct simulation of the system (\ref{eq:Ham}-\ref{eq:Liou}) by the
Monte Carlo wave-function (MCWF) method, which unravels the
corresponding Master equation in terms of individual quantum
trajectories. This approach takes into account the full particle and
cavity dynamics, with the cavity decay accounted for by quantum jumps,
and can be practically pursued up to a few particles with the help of
a new simulation framework \cite{vukics07}. The second approach
analogous to standard Bose Hubbard models is based on a
second-quantised form of the Hamiltonian (\ref{eq:Ham}): \(\int
dx\,\Psi^\dag(x)\,H\,\Psi(x)\) where the field operator \(\Psi(x)\) is
restricted to the lowest vibrational band of the lattice.

To obtain the smallest possible system useful for studying
self-organisation, we restrict the dynamics to only one lattice
wavelength, that is, two lattice sites (cf \Fref{fig:Scheme}) with
periodic boundary condition. This is the smallest system which can
seize the difference between the configurations described above and
contains all the essential physics. 

\newcommand\rml{\mathrm{l}}
\newcommand\rmr{\mathrm{r}}
\newcommand\ketl{\ket{\rml}}
\newcommand\ketr{\ket{\rmr}}
\newcommand\bopl{b_\rml}
\newcommand\bopr{b_\rmr}

With two lattice sites the lowest vibrational band constitutes a
two-dimensional Hilbert space, for which the localised Wannier basis
with state \(\ketl\) localised at the left and \(\ketr\) at the
right lattice site can be used. (In the MCWFS there are several left
and right states corresponding to a large number of vibrational
bands.) Hence \(\Psi(x)=\braket{x}{\rml}\,\bopl+\braket{x}{\rmr}\,\bopr\),
where \(\bopl\) and \(\bopr\) are the corresponding bosonic annihilation
operators. Putting the restricted field operator back into the
second-quantised Hamiltonian we obtain the Bose-Hubbard type
Hamiltonian:
\begin{equation}
\label{eq:HamBH}
H_\mathrm{BH}=J\lp\bopl^\dag\bopr+\bopr^\dag\bopl\rp-\lp\DC-NU_0\rp a^\dag
a+\tilde{J}\lp\bopl^\dag\bopl-\bopr^\dag\bopr\rp\lp a^\dag+a\rp,
\end{equation}
where \(J\equiv\bra{\rml}(p^2/(2\mu)+V_0\,\sin^2(Kx))\ketr\) and
\(\tilde{J}/(\mathrm{sign}(U_0)\sqrt{U_0\,V_0})\equiv\bra{\rml}\sin(Kx)\ketl=-\bra{\rmr}\sin(Kx)\ketr\).

The dynamics of particles in cavities as described by such
Hamiltonians in system configurations different from the one
investigated here has been discussed in
Refs.~\cite{maschler05a,maschler05b,larson06}. A very attractive
feature of Hamiltonian (\ref{eq:HamBH}) is that it is simple enough so
that together with the Liouvillean (\ref{eq:Liou}) the full
time-dependent Master equation can be solved even for several
particles.

\section{Single-particle dynamics}
We first consider the dynamics of a single particle initially prepared
in one of the localised states (say, \(\ketr\)). Without coupling to
the cavity (\(U_0=0\)) the particle moves unperturbed in the lattice
via tunnelling, which, in the case of two sites corresponds to an
oscillation between states \(\ketr\) and \(\ketl\). This can be
monitored via the expectation value \(\avr{Kx}\), which, as displayed
in \Fref{fig:OneParticle}(a) (red line), oscillates accordingly
between \(\pm\pi/2\) (cf also \Fref{fig:Scheme}).

\newcommand{\omrec}{\omega_{\mathrm{rec}}}

\begin{figure}
\centering
\begin{tabular}{c}
\includegraphics[width=2.75in,angle=-90]{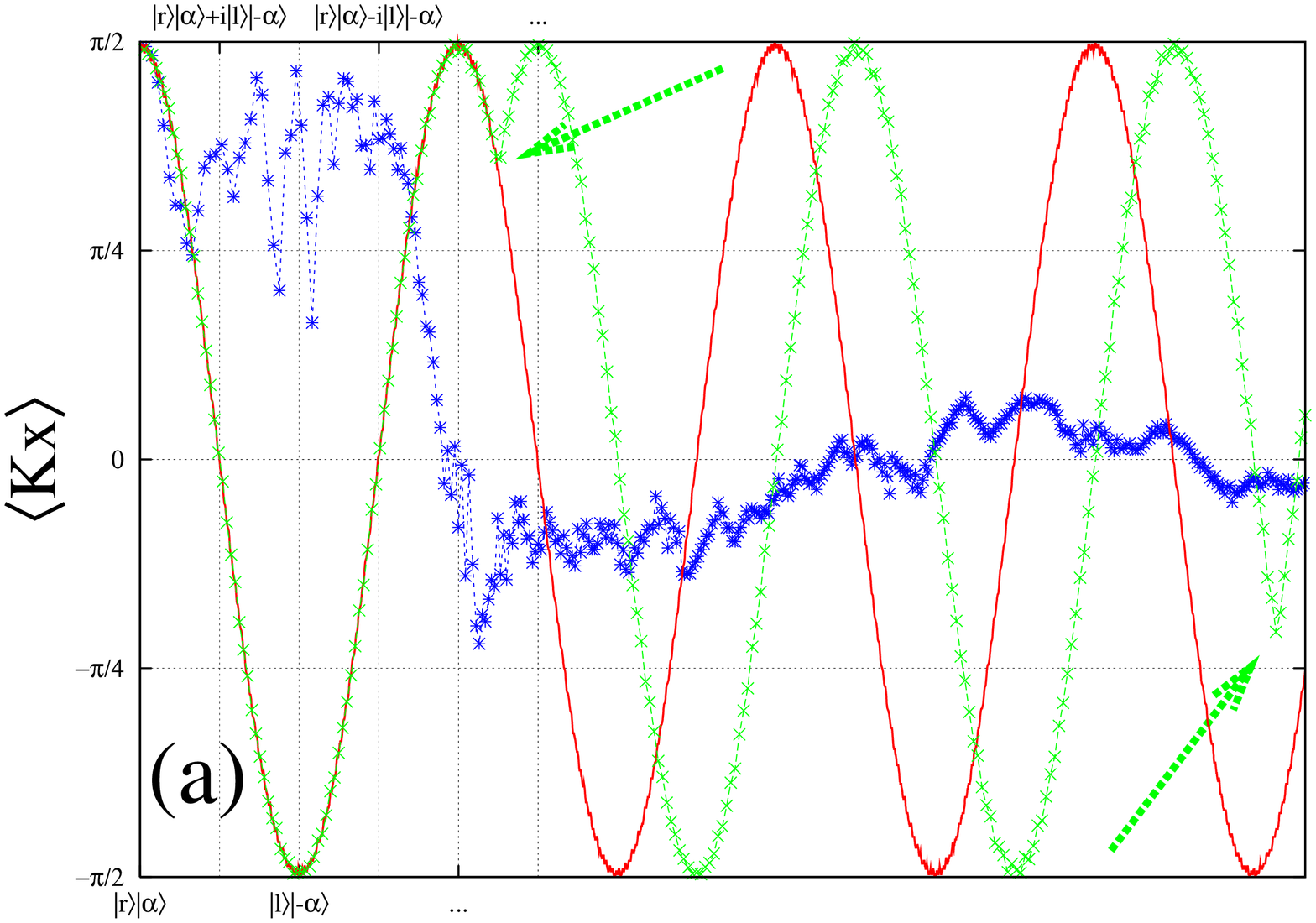}\\
\includegraphics[width=2.75in,angle=-90]{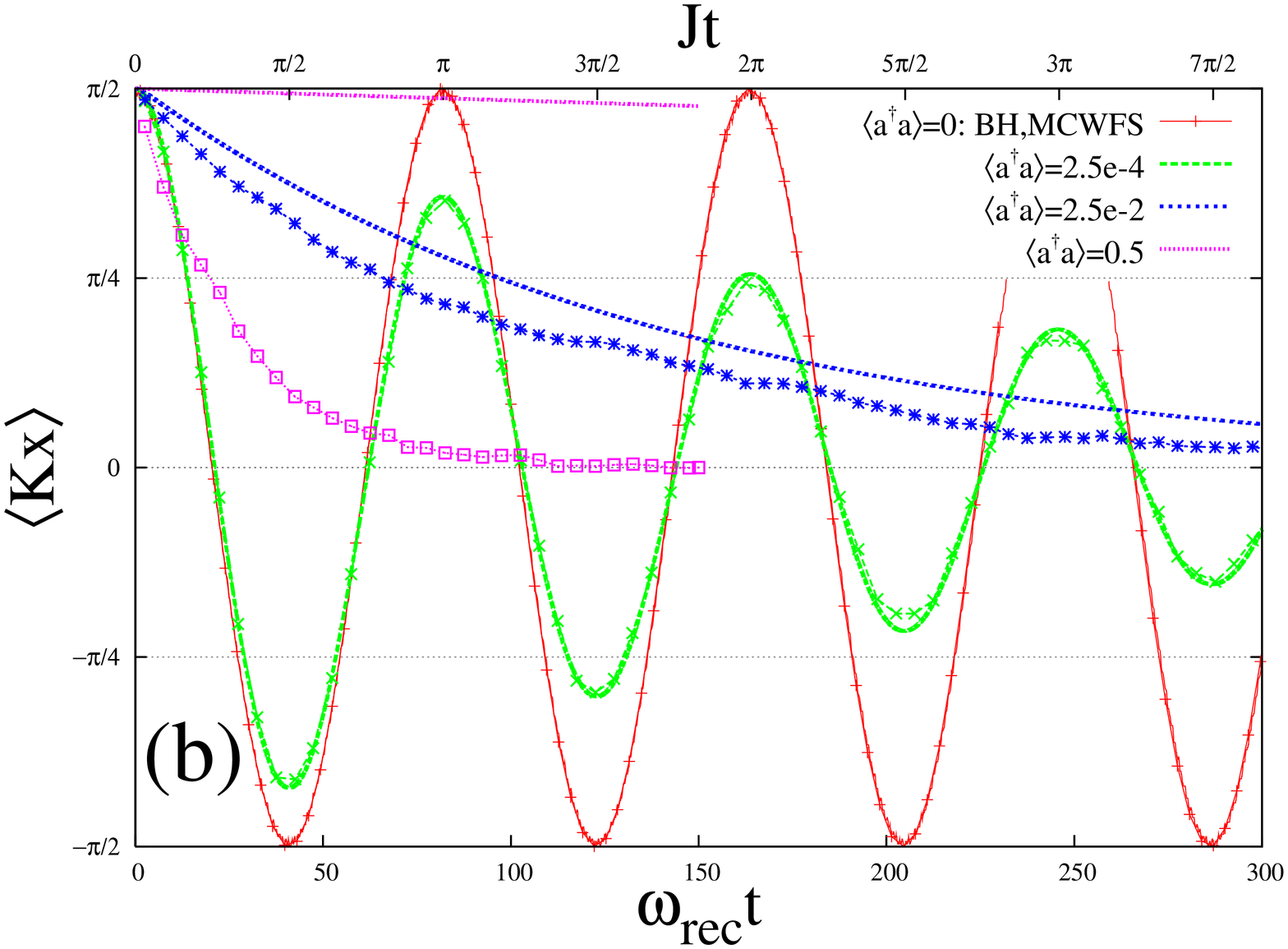}
\end{tabular}
\caption{Simulated data for a single particle in the lattice-cavity
  system. Parameters: \(V_0=-10\,\omrec\), \(\kappa=10\,\omrec\),
  \(\DC-U_0=-\kappa\), with the recoil frequency \(\omrec\equiv\hbar
  K^2/(2\mu)\). The colour code: red corresponds to \(U_0=0\), green
  to \(U_0=-0.005\omrec\), blue to \(U_0=-0.5\omrec\), and magenta to
  \(U_0=-10\,\omrec\); with maximal cavity photon numbers amounting to
  \(0\), \(2.5\eex{-4}\), \(2.5\eex{-2}\), and \(0.5\),
  respectively. (a) Example single MCWF trajectories. On the green
  trajectory green arrows mark the instants of cavity decays (photon
  escapes): the resulting jump in the tunnelling oscillation's phase
  is visible. (b) Ensemble average data --- Lines without points:
  solution of the time-dependent Master equation based on the
  Bose-Hubbard Hamiltonian (\ref{eq:HamBH}). Lines with points:
  ensemble average of several MCWF trajectories.}
\label{fig:OneParticle}
\end{figure}

This simple behaviour is significantly modified in the presence of
even a weakly coupled cavity, \(U_0\neq0\). Now the particle scatters
photons from the lattice field into the cavity mode, depending on its
state. Photons can decay according to the Liouvillean (\ref{eq:Liou})
and allow to monitor the particle motion. The decay of a cavity photon
can be modelled by a quantum jump, which is mathematically described
by the application of the cavity field operator \(a\) on the state
vector of the system. This, in turn, changes the whole particle-field
wave function and thus gives feedback on the particle localisation.

When the coupling is weak enough, the field in the cavity will be
small, and the contribution of the last term of Hamiltonian
(\ref{eq:Ham}) to the potential felt by the particle (second term in
the same Hamiltonian) is negligible, so that it still makes sense to
define the localised particle states solely from the lattice
potential.

We assume that these states are \emph{well} localised. When a
point-like particle is placed into the lattice at position \(x\) in
steady state it radiates a coherent field \(\ket{\alpha(x)}\) into the
cavity where the amplitude is determined by the Liouvillean dynamics
(\ref{eq:Ham}-\ref{eq:Liou}) and reads
\begin{equation}
\alpha(x)=\frac{\sqrt{U_0V_0}}{U_0-\DC-i\kappa}\sin(Kx)=
\frac{\sqrt{U_0V_0}}\kappa\frac1{1-i}\sin(Kx),
\end{equation}
where the second equality holds under the resonance condition
\(\DC-NU_0=-\kappa\) (\(N\) is the particle number), to which we
restrict ourselves in the following. This makes that the cavity field
increases monotonically with increasing coupling.

Accordingly, in state \(\ketr\) the particle will radiate an
approximately coherent state \(\ket\alpha\), while in state \(\ketl\)
a coherent state with opposite phase \(\ket{-\alpha}\), where
\(\alpha=\alpha(x=\pi/(2K))\).

If we assume that tunnelling is much slower than cavity field
evolution, then the latter will follow adiabatically the
former. Without cavity jumps the system evolves coherently and since
the back action of the cavity field on the particle motion is
negligible by our assumption, this evolution amounts to an oscillation
between states \(\ket{\rmr,\alpha}\) and \(\ket{\rml,-\alpha}\), hence at a
given time instant \(t\) the overall state of the particle-cavity
system reads approximately
\begin{equation}
\ket{\Psi(t)}=\cos(Jt)\ket{\rmr,\alpha}+i\sin(Jt)\ket{\rml,-\alpha}.
\end{equation}
Now imagine that at time \(t\) a jump happens: Immediately after the
jump the state of the system reads
\begin{equation}
\ket{\Psi'(t)}\propto a\ket{\Psi(t)}\propto
\cos(Jt)\ket{\rmr,\alpha}-i\sin(Jt)\ket{\rml,-\alpha},
\end{equation}
that is, the cavity jump is reflected back onto the particle motion and
results in a jump of the phase of the tunnelling oscillation. 

This behaviour is verified by the simulations, an example trajectory
is displayed in \Fref{fig:OneParticle}(a) (green line). Here the
parameters were chosen such that the maximal expectation value of the
cavity photon number is only \(2.5\eex{-4}\) --- this maximum is
achieved when the particle is prepared perfectly localised at a
lattice site.

The jump is a stochastic event and in ensemble average the jumps of
the phases on individual trajectories result in a dephasing and hence
damping of the oscillation. This behaviour, as displayed in
\Fref{fig:OneParticle}(b) (green lines) is verified by both the MCWFS
and the simulation of the time-dependent Master equation based on the
Bose-Hubbard Hamiltonian (\ref{eq:HamBH}). In this regime of very low
cavity photon number, the correspondence between the two models is
very good. Increasing the photon number results in several jumps
happening in one tunnelling cycle: in ensemble average this
corresponds to the over-damped regime of the tunnelling oscillation
(cf \Fref{fig:OneParticle}(a-b) blue line).

The above picture of the dynamics on one Monte Carlo trajectory is not
valid in the regime of stronger coupling where the photon number is
higher. Here the cavity field modifies significantly the potential
felt by the particle and hence the states \(\ketr\) and \(\ketl\)
defined solely by the lattice potential lose their significance
because many other particle spatial states enter the dynamics. Cavity
decays are much more frequent, and the stronger field fluctuations are
reflected back onto the potential. Accordingly, as we observe in
\Fref{fig:OneParticle} (magenta line), the Bose-Hubbard approach being
based on those two states breaks down in this regime.

\begin{figure}
\centering
\includegraphics[angle=-90,width=4in]{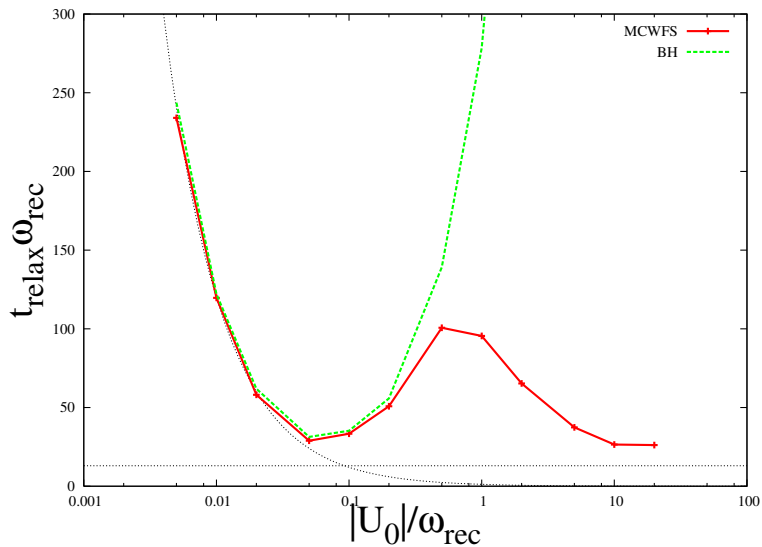}
\caption{Relaxation time of the damping of the tunnelling oscillation
  simulated with the MCWF and Bose-Hubbard approach. Same parameters
  as above. The dotted hyperbole has been fit on the weak-coupling
  part of the data and is proportional to the cavity photon count
  rate. The dotted horizontal line is the time scale of tunnelling
  with no coupling.}
\label{fig:relaxTime}
\end{figure}

The steady state of the dynamics is, quite independently of the
coupling, the mixed state
\(\rho_{\mathrm{ss}}\approx(\ket{\rmr,\alpha}\bra{\rmr,\alpha}+
\ket{\rml,-\alpha}\bra{\rml,-\alpha})/2\).
As displayed in \Fref{fig:relaxTime}, the relaxation time to this
state, that is, the time scale of the damping of the tunnelling
oscillation exhibits a non-trivial behaviour. For moderate coupling,
according to the discussion above, this is composed of two time
scales: the time scale of cavity photon decay, and that of
tunnelling. The former being inversely proportional to the photon
number gets faster with increasing coupling.

On the other hand the latter, for far less obvious reasons, \emph{gets
slower}. The discussion of the exact mechanism of this slowing down of
the tunnelling is not in the scope of the present paper, and must be a
subject of further study. Here we merely note that this behaviour is
verified by both the MCWF and the Bose-Hubbard approach, and can
phenomenologically be considered as an increase of the particle's
effective mass. It is then analogous with the effective mass of
electrons in crystals: there, when tunnelling between lattice sites,
not only the electron has to tunnel, but a quasi particle called
polaron composed of the electron and phonons. (The polaron problem has
huge literature, \cite{landau46} is one of the earliest references,
for a quite comprehensive review see eg \cite{mahan}). Here, roughly
speaking, when the particle tunnels from \(\ket{r}\) to \(\ket{l}\),
the cavity field also has to \emph{tunnel} from \(\ket{\alpha}\) to
\(\ket{-\alpha}\), which is inhibited by increasing \(\alpha\). Note
that here it is a single EM mode that generates the effective mass,
instead of several phonon modes as in the above polaron analogue. Here
we see an example of the combination of optical lattices and CQED
being able to reproduce a wider range of solid state physics phenomena
(in particular the existence of phonons) than optical lattices alone.

The combination of one accelerating and one slowing time scale results
in the minimum behaviour of the relaxation time in
\Fref{fig:relaxTime} around \(U_0=-0.05\,\omrec\). Ultimately, with
high enough coupling the cavity-generated potential starts to dominate
the lattice potential, in which regime strong fluctuations and
self-trapping lead to fast relaxation as observed in the MCWFS (red
line \(\abs{U_0}\gtrsim\omrec\)). Obviously, the Bose-Hubbard approach
is unable to reproduce the behaviour in this regime.

\newcommand\nl{n_\rml}
\newcommand\nr{n_\rmr}

\section{Two-particle dynamics}
Having understood the dissipative quantum dynamics of a single
particle in our lattice-cavity system, we now turn to the case of two
particles. Two particles on two lattice sites with periodic boundary
conditions is the minimal system that can exhibit the difference
between the configurations described above for self-organisation. In
the Bose-Hubbard approach where there is only one state at each
lattice site the Hilbert space for the particles is spanned by only
three states: \(\ket{1,1}\equiv\ket{0}\) --- the Mott insulator (MI)
state, which corresponds to the homogeneous distribution or
unorganised configuration ---, and \(\ket{2,0}\equiv\ket{-}\) and
\(\ket{0,2}\equiv\ket{+}\) corresponding to the two organised
configurations. \(\ket{0}\) scatters no field (and no photons) into
the cavity due to destructive interference between the fields
scattered by the two particles, while \(\ket{-}\) and \(\ket{+}\)
scatter \(\ket{-2\alpha}\) and \(\ket{2\alpha}\), respectively, the
factor 2 being due to constructive interference.  The difference
between the two configurations can be monitored via the density
correlation \(\avr{\nl\nr}\), which is 1 for the MI state and 0 in the
subspace spanned by \(\ket\pm\).

When the particles are initially in the MI state, then for \(U_0=0\)
at time \(t\) the particle state is
\begin{equation}
\ket{\Psi(t)}=\cos(2Jt)\ket0+\frac i{\sqrt2}\sin(2Jt)\lp\ket-+\ket+\rp.
\end{equation}
With \(U_0\neq0\) under the simplifying conditions we discussed above
for the single particle case we have for the joint system
\begin{equation}
\label{eq:tunnelling}
\ket{\Psi(t)}=\cos(2Jt)\ket{0,0}+
\frac i{\sqrt2}\sin(2Jt)\lp\ket{-,-2\alpha}+\ket{+,2\alpha}\rp.
\end{equation}

\begin{figure}
\centering
\includegraphics[width=2.75in,angle=-90]{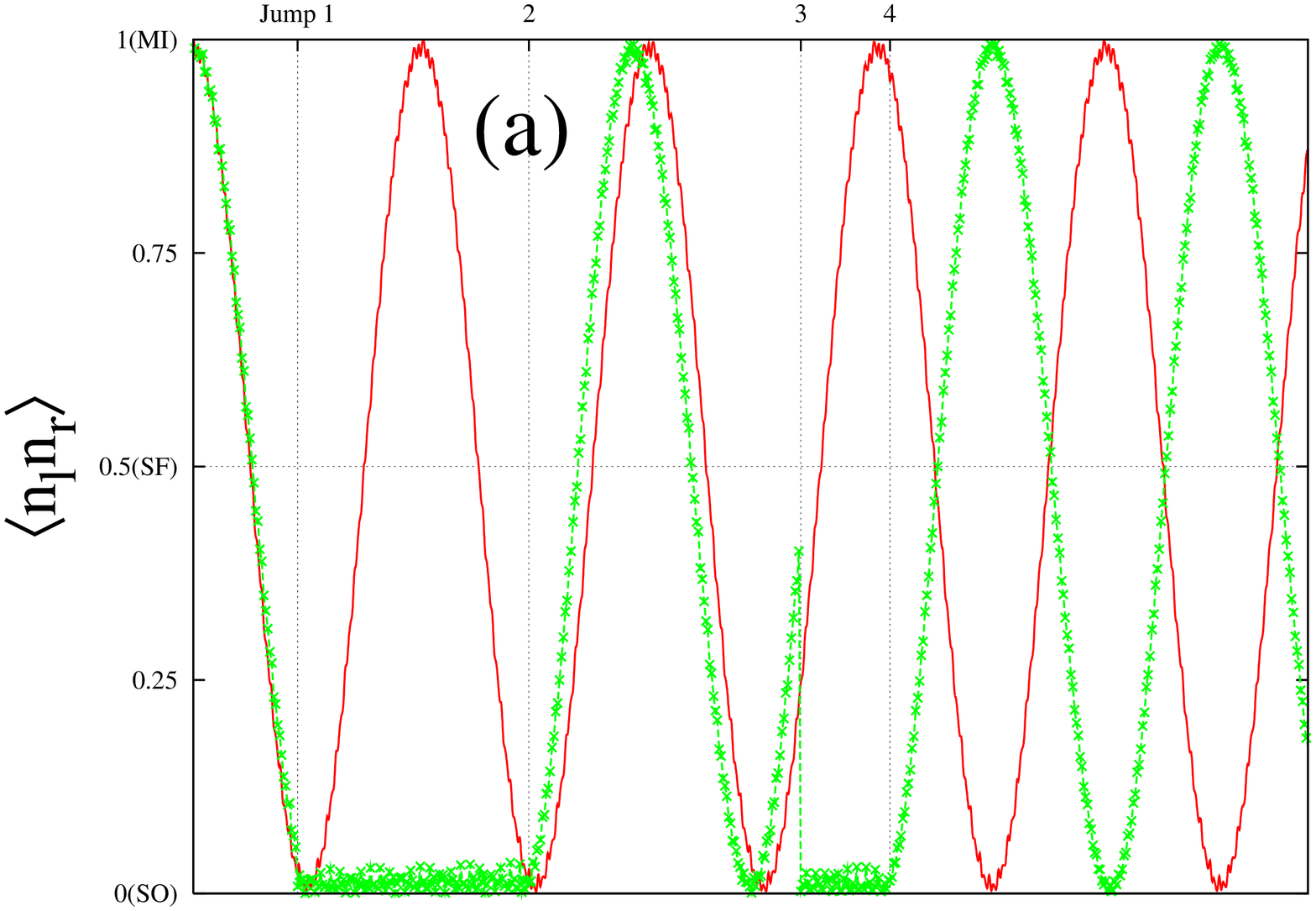}\\
\includegraphics[width=2.75in,angle=-90]{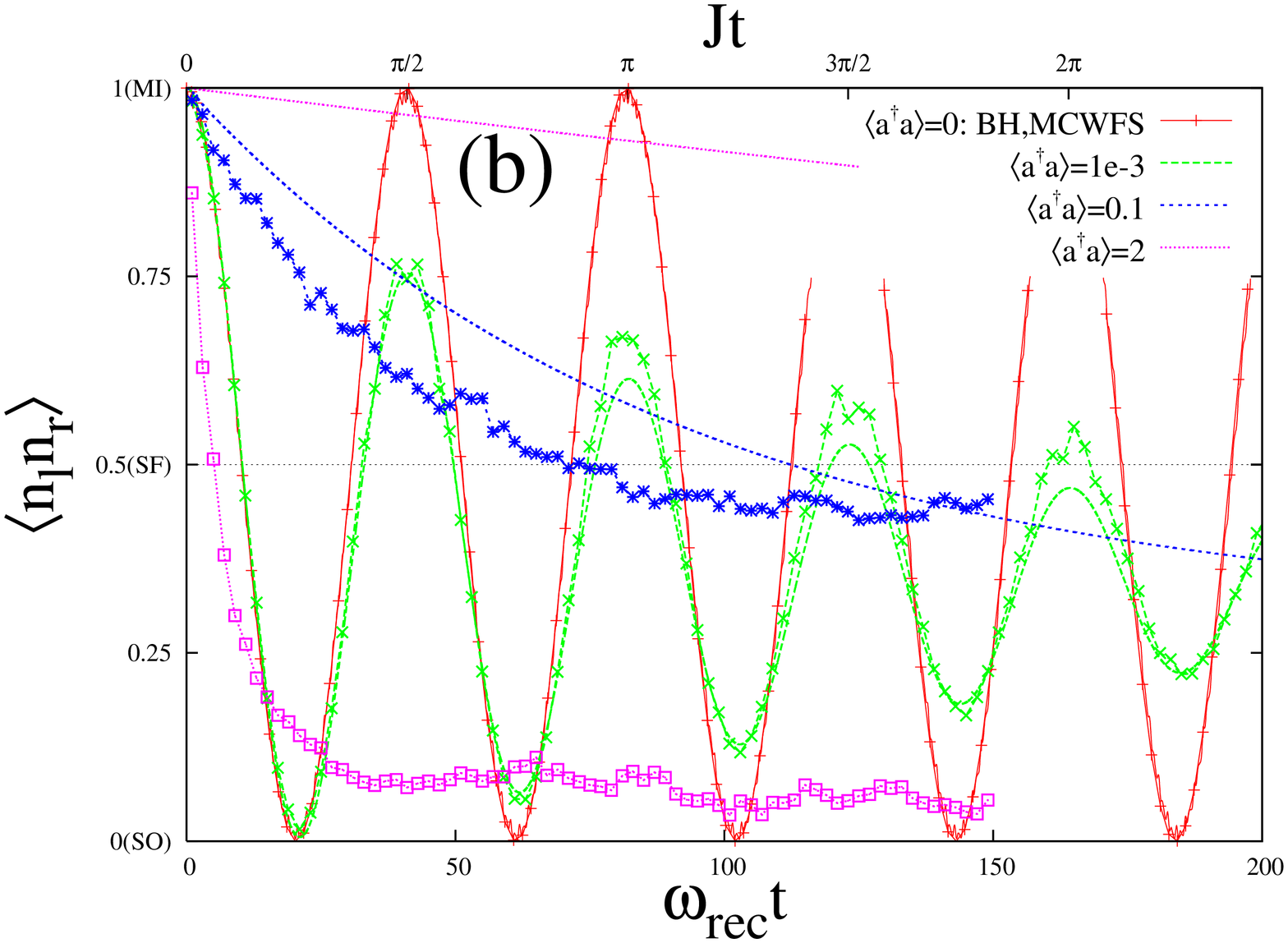}
\caption{Simulated data for two particles in the lattice-cavity system
  with same parameters as above. The colours correspond to the same
  values of \(U_0\), here with maximal cavity photon numbers amounting
  to \(0\), \(10^{-3}\), \(0.1\), and \(2\). (a) Example single MCWF
  trajectories --- the vertical dotted lines mark the time instants of
  cavity decays on the trajectory plotted in green. (b) Ensemble
  average data --- lines without points: data stemming from the
  Bose-Hubbard approach; lines with points: MCWF approach.}
\label{fig:TwoParticles}
\end{figure}

If a jump happens in this state (application of \(a\)), then the state
immediately after the jump reads
\begin{equation}
\label{eq:Dark}
\ket{\Psi'(t)}\propto a\ket{\Psi(t)}\propto
\ket{-,-2\alpha}-\ket{+,2\alpha}.
\end{equation}
There are two points worth noting here: Firstly, the quantum jump in
the photon number erases all information about the phase of the
tunnelling oscillation in the particle Hilbert space. Secondly, after
the escape of one single photon from the cavity tunnelling stops
immediately.
Indeed, in the state (\ref{eq:Dark})
both \(\ket{-,-2\alpha}\) and \(\ket{+,2\alpha}\) tunnels to
\(\ket{0,0}\) (note that we assume again the cavity field following
adiabatically the tunnelling), but their phases are opposite, and
hence the two paths interfere destructively. A second jump at \(t'>t\),
however, makes the phases match again, and puts the state
\(\ket{\Psi''(t')}\propto
a\ket{\Psi'(t)}\propto\ket{-,-2\alpha}+\ket{+,2\alpha}\) back into the
tunnelling cycle (\ref{eq:tunnelling}). An example MCWF trajectory
exhibiting this behaviour is plotted in \Fref{fig:TwoParticles}(a)
(green line). We observe that a quantum jump brings the system into
the state (\ref{eq:Dark}), signalled by \(\avr{\nl\nr}=0\), and it
remains there until the next jump, when it starts to oscillate anew.

In ensemble average these stochastic ``dark'' periods of the
tunnelling oscillation lead to damping just as in the single-particle
case. The final steady state is always a mixture
\begin{equation}
\label{eq:SS}
\rho_{\mathrm{ss}}=w\ket{0,0}\bra{0,0}
+\frac{1-w}2\lp\ket{-,-2\alpha}\bra{-,-2\alpha}
+\ket{+,2\alpha}\bra{+,2\alpha}\rp.
\end{equation}
At this point it becomes clear that any mean-field description of this
system is bound to fail: a mean cavity field description would
prohibit the possibility of a coherent superposition of different
particle configurations radiating different fields as in
(\ref{eq:tunnelling}), which is essential for the onset of the
dynamics from a homogeneous zero-temperature initial condition (see
also \cite{maschler07}). On the other hand, a particle mean field cannot
capture the difference between states (\ref{eq:SS}) with different
\(w\), because this appears only in the density \emph{correlation}.

\begin{figure}
\centering
\includegraphics[width=2.75in,angle=-90]{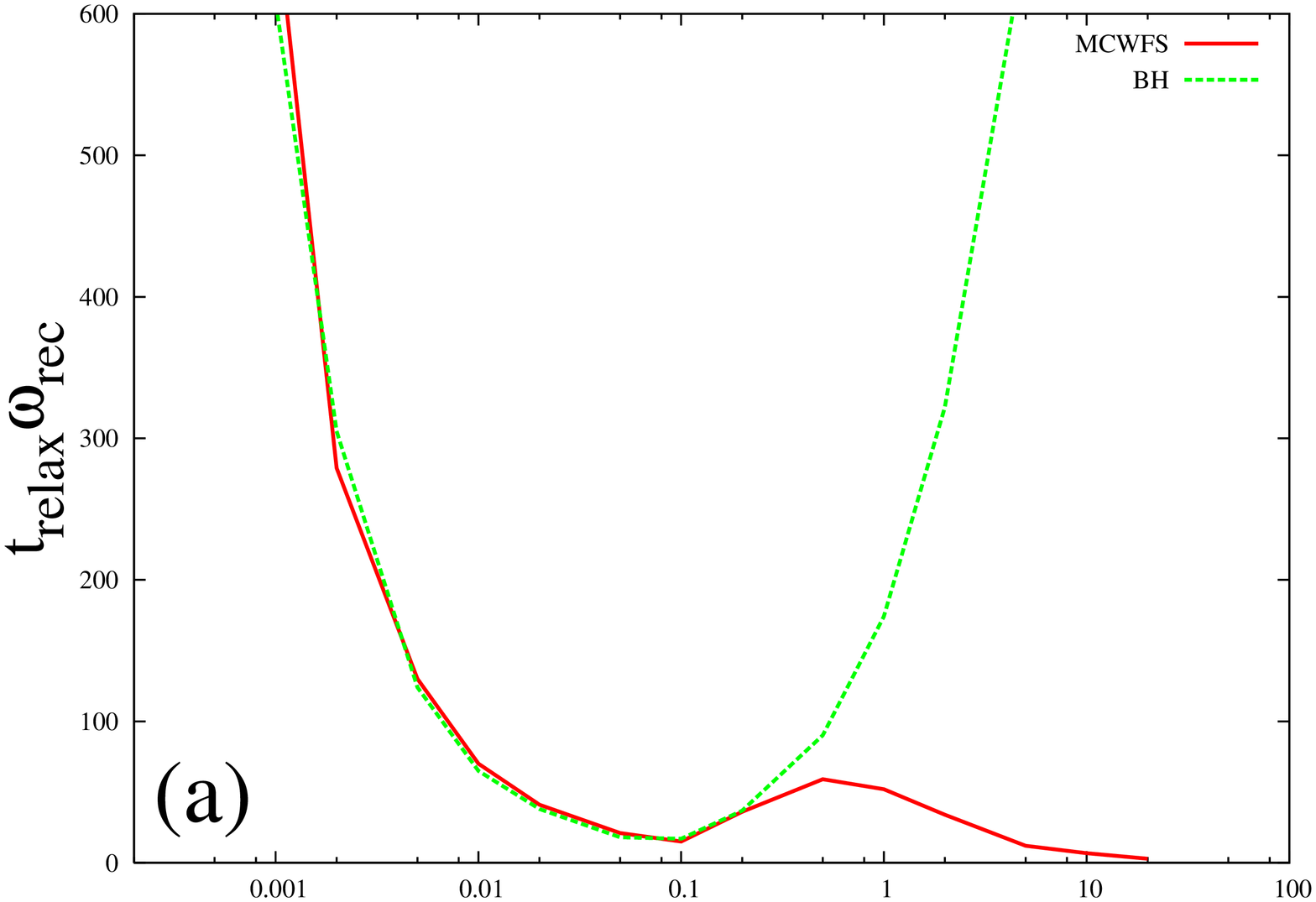}\\
\includegraphics[width=2.75in,angle=-90]{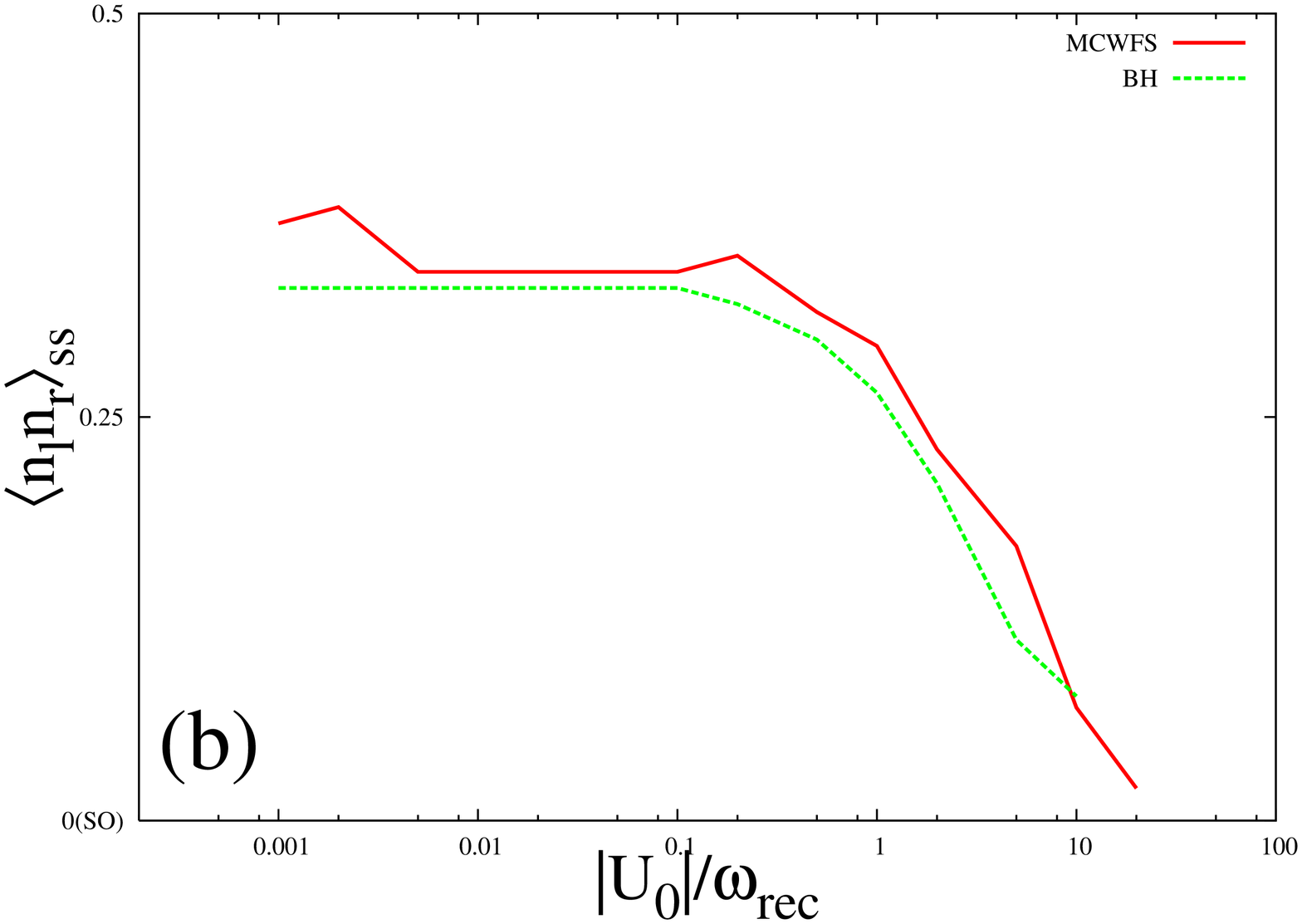}
\caption{(a) Relaxation time for two particles and (b) weight of the
  MI component in the steady-state mixture (\ref{eq:SS}) ---
  \(\avr{\nl\nr}_{\mathrm{ss}}=w\) --- as functions of the
  coupling. Same parameters as above.}
\label{fig:relax}
\end{figure}

As displayed in \Fref{fig:TwoParticles}(b), our two approaches for
simulating the damping agree well in the regime of moderate
coupling. However, just as in the single-particle case, strong
coupling --- in the regime where the cavity-generated potential
dominates the lattice one --- results in extremely quick damping,
which cannot be reproduced by the Bose-Hubbard approach. Here the
relaxation time exhibits the same behaviour as we have seen in the
single-particle case (cf \Fref{fig:relax}(a)): with increasing
coupling it has a minimum, after which the Bose-Hubbard approach gives
a monotonic increase of the relaxation time, while the MCWFS gives a
peak, and for even stronger coupling very quick relaxation. For the
interpretation of this behaviour the same discussion applies as above
for the single-particle case.

Increasing coupling results in a decrease of the weight \(w\) of the
MI component in steady state, cf \Fref{fig:relax}(b), and with strong
enough coupling (\(\abs{U_0}\gtrsim10\,\omrec\) in our case) the
steady state is confined solely into the self-organised subspace. This
proves our initial assertion that even when the system is started from
a \(T=0\) homogeneous state (here the MI state) self-organisation can
occur. On the same Figure we also see that this is confirmed by both
approaches, only the relaxation time of the process is predicted
wrongly by the Bose-Hubbard approach for strong coupling.

\begin{figure}
\centering
\includegraphics[width=2.75in,angle=-90]{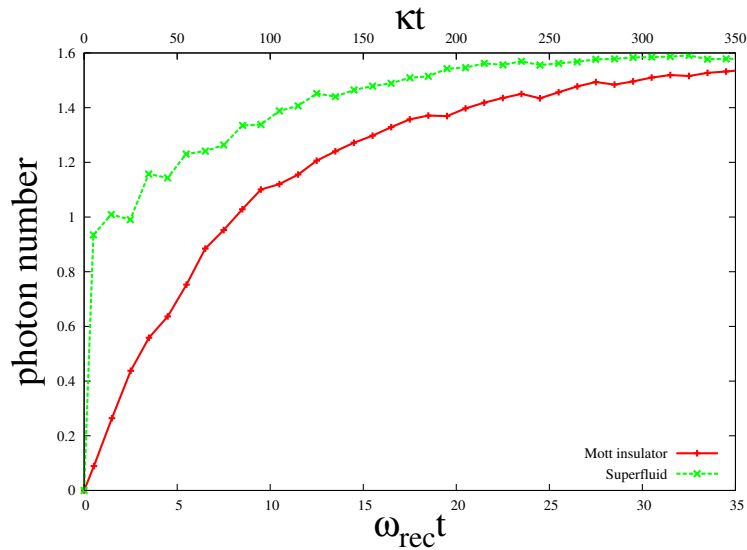}
\caption{Initial build-up of the cavity photon number with the
  particles prepared initially in the Mott insulator or superfluid
  state. \(U_0=-10\omrec\), other parameters are the same as above.}
\label{fig:PhotNum}
\end{figure}

It is easy to see that starting the system from the superfluid (SF)
state
\begin{equation}
\ket{\mathrm{SF}}\propto\ket0+\frac1{\sqrt2}\lp\ket-+\ket+\rp
\end{equation}
instead of the MI as above does not change the steady state since
already the first quantum jump erases the information about the
initial condition completely. The process of relaxation will, however,
be different. This process can be monitored by a time resolved
analysis of the intensity escaping the cavity, which is proportional
to the cavity photon number: an example is displayed in
\Fref{fig:PhotNum}. Here, we are in the over-damped regime of the
tunnelling oscillation. When prepared in the MI state, the particles
do not scatter initially, and the buildup of the cavity field occurs
on the time scale of the self-organisation process. With the SF
initial state, on the other hand, some field is built up almost
instantly (on the time scale of \(\kappa^{-1}\)), because in the SF
the states \(\ket\pm\) have finite weight; while the rest of the field
is built up on the longer time scale.

\section{Conclusions}
In summary, we have seen that coupling an optical lattice gas at
\(T=0\) to a cavity induces an irreversible reorganisation of the
particles, a process which can be monitored directly in an experiment by
the time-resolved analysis of the intensity escaping the cavity. We
have shown that no classical or mean-field description of either the
particles or the cavity field can account for the phenomenon. For strong
enough coupling the process leads to a fast self-organisation of the
particles, a phase in which they occupy every second site in the lattice,
and scatter superradiantly into the cavity. An important conclusion of
the work is that while an effective low-dimensional Bose-Hubbard type
model cannot reproduce the time evolution in the strong-coupling
regime as observed in more detailed simulations, it can still predict
the steady state remarkably well. This model can therefore be used in
the future for a high number of particles to study possible quantum phase
transitions occurring in the steady state of this dissipative system.

\section*{References}

\bibliographystyle{unsrt}
\bibliography{quantorg}

\end{document}